\documentclass{eptcs}

\usepackage{colortbl}
\usepackage{times}
\usepackage{pifont}
\normalfont
\usepackage[T1]{fontenc}
\usepackage{versions}
\usepackage[tight]{subfigure}

\usepackage{tikz}
\usetikzlibrary{shapes.multipart}
\usetikzlibrary{automata}
\usetikzlibrary{positioning}
\usepgflibrary{shapes.gates.logic.US}
\usetikzlibrary{shapes.gates.logic.US}
\usetikzlibrary{shadows}
\usetikzlibrary{arrows}
\usetikzlibrary{trees}

\usepackage{xspace,mathpartir}
\usepackage[draft,inline,index,marginclue,nomargin]{fixme}
\usepackage{graphicx, xspace, amssymb, latexsym, amscd}
\usepackage{amsmath, amssymb, amsbsy}
\usepackage{mathtools}
\definecolor{Blue}{rgb}{0.3,0.3,0.9}

\newcommand{\Sect}[1]{Section~\ref{sec:#1}\xspace}

\newcommand{\embox}[1]{\mbox{\textit{#1}}}

\newcommand{\df}{\stackrel{{\tiny df}}{=}}
\newcommand{\st}{\;\cdot\;}

\newcommand{\refdef}[1]{Definition~\ref{def:#1}\xspace}

\newtheorem{definition}{Definition}
\newtheorem{proposition}{Proposition}
\newtheorem{theorem}{Theorem}
\newenvironment{proof}{\noindent\textbf{Proof:}}{}

\newcommand{\other}[1]{\ensuremath{\overline{#1}}\xspace}
\newcommand{\conflictsymbol}{\maltese}
\newcommand{\conflicts}{\;\conflictsymbol\;}

\newcommand{\sync}[1]{\|_{{\small  #1}}}
\newcommand{\acts}[1]{\embox{acts}(#1)}
\newcommand{\nxt}[1]{\embox{next}(#1)}

\newcommand{\goesto}[1]{\xrightarrow{#1}}

\newcommand{\contractF}{\embox{contract}}
\newcommand{\contract}[1]{\contractF(#1)}
\newcommand{\mexcl}{\bowtie}

\newcommand{\pp}{{\mathcal P}}

\newcommand{\oo}{{\mathcal O}}
\newcommand{\ff}{{\mathcal F}}

\newcommand{\OPpp}[2]{\ensuremath{\pp_{#1}(#2)}\xspace}

\newcommand{\OPff}[2]{\ensuremath{\ff_{#1}(#2)}\xspace}
\newcommand{\OPoo}[2]{\ensuremath{\oo_{#1}(#2)}\xspace}

\newcommand{\viable}[3]{\embox{viable}_{#1}(#2,#3)}

\newcommand{\satSingle}[3]{#1 \vdash_{#2} #3}

\newcommand{\satP}[2]{\embox{sat}_{#1}^P(#2)}
\newcommand{\satO}[2]{\embox{sat}_{#1}^O(#2)}

\newcommand{\satA}[3]{\embox{sat}^{#1}_{#2}(#3)}
\newcommand{\sat}[2]{\embox{sat}_{#1}(#2)}
\newcommand{\semantics}[1]{[\![ #1 ]\!]}
\newcommand{\bi}[2]{\embox{breachIncapable}_{#1}(#2)}
\newcommand{\CA}{{\mathcal C}{\mathcal A}}

\newcommand{\LEQcontract}{\sqsubseteq}

\newcommand{\Clause}{\mbox{Clause}}

\makeatletter
\newcommand*{\textol}[1]{$\overline{\hbox{#1}}\m@th$}
\makeatother

\newenvironment{myeqnarray}{\[\begin{array}{rcl}&&\\[-2.3em]}{\\&&\\[-2.3em]\end{array}\]}

\newcommand{\etal}{\emph{et al.}\xspace}

\excludeversion{TR}
\includeversion{PAPER}

\begin{document}

\title{Contracts for Interacting Two-Party Systems}
\def\titlerunning{Contracts for Interacting Two-Party Systems}

\author{
  Gordon J. Pace
  \institute{Department of Computer Science,\\University of Malta}
  \email{gordon.pace@um.edu.mt}
  \and 
  Fernando Schapachnik\thanks{Partially founded by UBACyT 20020100200103.}
  \institute{Departamento de Computaci\'on, FCEyN,\\ Universidad de Buenos Aires, Buenos Aires, Argentina}
  \email{fschapachnik@dc.uba.ar}
}
\def\authorrunning{Fernando Schapachnik, Gordon J. Pace}
\maketitle

\begin{abstract}
This article deals with the interrelation of deontic operators in contracts -- an aspect often neglected when considering only one of the involved parties. On top of an automata-based semantics we formalise the onuses that obligations, permissions and prohibitions on one party impose on the other. Such formalisation allows for a clean notion of contract strictness and a derived notion of contract conflict that is enriched with issues arising from party interdependence.
\end{abstract}

\section{Introduction}
\label{sec:intro}

Deontic modalities such as permission and obligation have been debated exhaustively in the literature, and various formalisms exist with different interpretations and axiomatisation of deontic notions. With few exceptions, the modalities are usually presented in an impersonal manner, refering only to the subject of the modality. For instance, most formalisms enable reasoning about notions such as ``John is permitted to withdraw cash'' and ``John is obliged to pay an annual credit card fee''. However, in a contractual setting, the behaviour involves interaction between the two parties the contract binds, and such statements about the ideal behaviour have both a notion of the subject and object of the action. For instance, in a contract between John and his bank, the clause ``John is permitted to withdraw cash'' is about both parties, and can be interpreted to mean that if John attempts to withdraw cash, then the bank will not refuse or hinder his action. Similarly, the clause ``John is obliged to pay an annual credit card fee'' places an obligation on John to perform an action with the bank as the object of the action, and (arguably) also places the onus on the bank to accept the payment. Interacting parties allow for both cooperation and interference between the parties in the actions they perform, and thus bring about an additional dimension to contract analysis. An interesting corollary to this view, is that permission can now be seen as a first class deontic modality. Typically seen as the dual of prohibition, violations of permissions have always proved difficult to formalise their violation, mainly since a branching logic analysis is required (if party $p$ were to perform $a$ then they would not be stopped from doing so). In an interacting two party system context, permission now takes a first class role, obliging the object of the modality to allow the subject to perform the action if they so desire. 

Although the work on deontic logic for interacting parties is not abundant, computer scientists have studied for various decades concurrent and synchronous composition, notions which embody precisely interaction from an action-based perspective. In \cite{PS_JURIX2011} we have presented work-in-progress on how synchrony can be applied in a contractual setting, using a formal automaton-based model of interacting two-party systems in which the parties synchronise over a set of actions. In this paper we extend the work presented there to deal with (i) absence of actions; (ii) mutually exclusive actions; (iii) conflicts. 

The rest of the paper is organised as follows. The next section formalises our notions of automata, deontic operators, contracts and contracts' strength, which allows us to show, in \Sect{conflicts} that some contracts cannot be satisfied at the same time and thus lead to a conflict.  Finally, in \Sect{related_work} we discuss related work, and conclude in \Sect{conclusions}.


\section{Regulated Two-Party Systems}
\label{sec:automata_semantics}

\subsection{An Automata-Based View}
To enable direct reasoning about contracts, one requires a model in which the two parties somehow interact to agree on which actions to perform. We use the notion of synchronous composition \cite{synchronous_composition} to model such behaviour. 
Furthermore, to be able to deal with concurrent obligations (for instance, one party being obliged to perform one action and the other being obliged to perform another), we adopt multi-action labels on transitions, since if we do not, it would be impossible not to violate a contract in which both parties have different obligations at the same time.

\begin{definition}
A multi-action automaton $S$ is a tuple $\langle \Sigma,\;Q,\;q0,\;\rightarrow\rangle$, where $\Sigma$ is the alphabet of actions, $Q$ is the set of states, $q0\in Q$ is the initial state and $\rightarrow \subseteq Q \times 2^\Sigma\times Q$ is the transition relation. 
We will write $q \goesto{A} q'$ for $(q,A,q')\in \rightarrow$, $\nxt{q}$ to be the set of target state and action set pairs of transitions outgoing from $q$ (defined to be $\{ (A,q') \mid q \goesto{A} q' \}$) and $\acts{q}$ to be the set of all action sets on the outgoing transitions from $q$ (defined to be $\{ A \mid \exists q' \st q \goesto{A} q'\}$). We say that an automaton is total, if for every $q\in Q$ and $A\subseteq \Sigma$, there is a $q'\in Q$ such that $q\goesto{A}q'$.

The synchronous composition of two automata $S_i=\langle Q_i,\;q0_i,\;\rightarrow_i\rangle$ for $i\in\{1,2\}$ (both with alphabet $\Sigma$) synchronising over alphabet $G$, written $S_1 \sync{G} S_2$, and is defined to be $\langle Q_1\times Q_2,\;(q0_1,q0_2),\rightarrow\rangle$, where $\rightarrow$ is the classical synchronous composition relation defined below:

\noindent\begin{minipage}{0.5\textwidth}
\begin{mathpar}
\inferrule*[right={$A \cap G = \emptyset$}]
{ q_1 \goesto{A}_1 q_1'}
{ (q_1,q_2) \goesto{A} (q_1',q_2) }
\end{mathpar}
\end{minipage}
\begin{minipage}{0.5\textwidth}
\begin{mathpar}
\inferrule*[right={$A \cap G = \emptyset$}]
{ q_2 \goesto{A}_2 q_2'}
{ (q_1,q_2) \goesto{A} (q_1,q_2') }
\end{mathpar}
\end{minipage}

\begin{mathpar}
\inferrule*[right={$A \cap G = B \cap G \neq \emptyset$}]
{ q_1 \goesto{A}_1 q_1',\;q_2 \goesto{B}_2 q_2' }
{ (q_1,q_2) \goesto{A\cup B} (q_1',q_2') }
\end{mathpar}
\end{definition}

\noindent We can now define contracts to be automata with each state tagged with the contract which will be in force at that point. The contracts will be able to refer to both presence and absence of an action. Given an alphabet of actions $\Sigma$, we write $\Sigma!$ to refer to the alphabet extended with actions preceded with an exclamation mark $!$ to denote their absence: $\Sigma! \df \Sigma \cup \{!a \mid a \in \Sigma \}$. We use variables $x$ and $y$ to range over $\Sigma!$. If $x$ is already an inverted action $x=!a$, then expression $!x$ is interpreted to be $a$.

Contract clauses are either (i) obligation clauses of the form $\OPoo{p}{a}$ or $\OPoo{p}{!a}$, which say that party $p$ is obliged to perform or not perform action $a$ respectively; or (ii) permission clauses which can be either of the form of $\OPpp{p}{a}$ or $\OPpp{p}{!a}$ (party $p$ is permitted to perform, or not perform action $a$ respectively).

\begin{definition}
A \emph{contract clause} over alphabet $\Sigma$ is structured as follows (where action $x\in \Sigma!$, party $p\in\{1,2\}$):\\
\centerline{$\Clause ::= \oo_p(x) \mid \pp_p(x)$}


\noindent A \emph{contract automaton} is a total and deterministic multi-action automaton \mbox{$S=\langle Q,\;q0,\;\rightarrow\rangle$,} together with a total function $\contractF \in Q \rightarrow 2^{\Clause}$ assigning a set of clauses to each state. We use $\CA$ to refer to the class of contract automata.

\noindent Two contract automata are said to be structurally isomorphic if they are structurally identical automata (they have the same set of states, initial state and transition relation) but may have different contract functions.
\end{definition}

\noindent Structurally isomorphic contract automata allow us to reason about the weakening or strengthening of a contract by changing the clauses in particular states but respecting the structure (and thus the temporal behaviour) of the contract, and will be used in various theorems in the rest of the paper. We can now define a regulated two-party system in terms of multi-action automata.

\begin{definition}\label{def:reg_two_party_system}
A \emph{regulated two-party system} synchronising over the set of actions $G$ is a tuple $R=\langle S_1, S_2\rangle^{\mathcal A}_G$, where $S_i=(\Sigma_i,Q_i,q0_i,\rightarrow_i)$ is a multi-action automaton specifying the behaviour of party $i$, and ${\mathcal A}$ is a contract automaton over alphabet $\Sigma_1\cup \Sigma_2$.

The behaviour of a regulated two-party system $R$, written $\semantics{R}$, is defined to be the automaton $(S_1 \sync{G} S_2) \sync{\Sigma} {\mathcal A}$. To make states in such systems more readable, we will write $((q_1,q_2),q_{\mathcal A})$ as $(q_1,q_2)_{q_{\mathcal A}}$.

A regulated two-party system is well-formed if $S_1 \sync{G} S_2$ never deadlocks: $\forall (q_1,q_2) \st \acts{q_1,q_2} \neq \emptyset$.
\end{definition}

\noindent In the rest of the paper we will assume that all systems are well-formed, i.e., do not deadlock. One way of guaranteeing this may be by having all system states provide a transition with the empty action.

\noindent Also note that the totality of the contract automaton guarantees that the system behaviour is not constrained, but simply acts to tag the states with the relevant contracts at each point in time.

\subsection{Contract Satisfaction}

Given a two-party system $(S_1,S_2)$, and a contract automaton ${\mathcal A}$, we can now define whether or not either party is violating the contract when a particular state is reached or a transition is taken. As we will see, a dual-view of violation, identifying \emph{both} bad states and bad transitions, is necessary in a deontic context. We will look at the different deontic operators and define the set of violations induced for each of them.  

\begin{definition}
Functions $O_p(q_{\mathcal A})$ and $F_p(q_{\mathcal A})$ give the set of actions respectively obliged to be performed and obliged not to be performed by party $p$. They are defined in terms of the contract clauses in the state. 
\begin{myeqnarray}
O_p(q_{\mathcal A}) & \df & \{ a \mid \oo_p(a) \in \contract{q_{\mathcal A}} \}\\
F_p(q_{\mathcal A}) & \df & \{ a \mid \oo_p(!a) \in \contract{q_{\mathcal A}} \}
\end{myeqnarray}
\noindent Action set $A$ is said to be viable for party $p$ in a contract automaton state $q_{\mathcal A}$, written $\viable{p}{q_{\mathcal A}}{A}$, if (i) all her obliged actions are included in $A$ but; (ii) no actions which the party is obliged not to perform are included $A$:
\begin{myeqnarray}
\viable{p}{q_{\mathcal A}}{A} & \df & 
  O_p(q_{\mathcal A}) \subseteq A  \land F_p(q_{\mathcal A}) \cap A = \emptyset
\end{myeqnarray}
\end{definition}



Since we would like to be able to place blame in the case of a violation, we parametrise contract satisfaction and violation by party. 

It is also worth noting that while obligation to perform an action, for instance, is violated in a transition which does not include the action, permission is violated by a state in which the opportunity to perform the permitted action is not present. The satisfaction predicate will thus be overloaded to be applicable to both states and transitions. The predicate $\satA{{\mathcal A}}{p}{X}$ will denote that the contract automaton ${\mathcal A}$, reaching state $X$ or traversing transition $X$, does not constitute a violation for party $p$. $X$ ranges over states and transitions in the composed system. When ${\mathcal A}$ is clear from the context, we simply write $\sat{p}{X}$. We start by defining separate satisfaction predicates for the deontic operators.

\paragraph{\textbf{Permission}.} If party $p$ is permitted to perform shared action $a$, then the other party $\other{p}$ must provide $p$ with at least one viable outgoing transition which contains $a$ but does not include any forbidden actions (that is, it is \emph{viable} for $p$). Permission to perform local actions cannot be violated. In the case of a single permission, this can be expressed as follows:
\[\begin{array}{l}
\satSingle{(q_1,q_2)_{q_{\mathcal A}}}{p}{\OPpp{p}{a}} \df true \\
\satSingle{(q_1,q_2)_{q_{\mathcal A}}}{\other{p}}{\OPpp{p}{a}} \df 
a \in G \implies
  \exists A \in \acts{q_{\other{p}}},\;A' \subseteq G^c \st 
    a \in A~\land~\viable{p}{q_{\mathcal A}}{A\cup A'}
\end{array}\]
Similarly, if party $p$ is permitted to not perform action $a$, then the other party $\other{p}$ must provide $p$ with at least one viable outgoing transition which does not include $a$ nor any forbidden action. Permission to perform local actions can never be violated. In the case of a single permission, this can be expressed as follows:
\[\begin{array}{l}
\satSingle{(q_1,q_2)_{q_{\mathcal A}}}{p}{\OPpp{p}{!a}} \df true \\
\satSingle{(q_1,q_2)_{q_{\mathcal A}}}{\other{p}}{\OPpp{p}{!a}} \df 
a \in G \implies
  \exists A \in \acts{q_{\other{p}}},\;A' \subseteq G^c \st 
    a \notin A~\land~\viable{p}{q_{\mathcal A}}{A\cup A'}
\end{array}\]
While actual obligation violations occur when an action is not performed, violations of a permission occur when no appropriate action is possible.
\begin{TR}
In this paper we give a semantics that tags as a violation a state in which one party is permitted to perform an action, while the other provides no way of actually doing so.
\end{TR}
For any other parameters, the permission is otherwise satisfied.

\noindent\textbf{Example:}
If $p$ is permitted to withdraw money from the bank, permitted not to deposit, obliged to pay the fee, and obliged not to steal (\OPpp{p}{w}, \OPpp{p}{!d}, \OPoo{p}{f}, \OPoo{p}{!s}), \other{p} should provide at least one transition that contains both a $w$ and an $f$ and contains neither a $d$ nor an $s$.


\noindent To combine all permissions in a state, we simply take the conjunction of all conditions:
\[
\satP{p}{(q_1,q_2)_{q_{\mathcal A}}} \df 
 \forall \OPpp{\other{p}}{x} \in q_{\mathcal A} \st 
    \satSingle{(q_1,q_2)_{q_{\mathcal A}}}{p}{\OPpp{\other{p}}{x}}
\]

\noindent All transitions are taken as satisfying the permission satisfaction function.

\paragraph{\textbf{Obligation}.} Obligation brings in constraints on both parties. Given that party $p$ is obliged to perform action $a$ in a state means that (i) party $p$ must include the action in any outgoing transition in the composed system in which it participates; and (ii) the other party $\other{p}$ must provide a viable synchronisation action set which, together with other asynchronous actions performed by $p$, allows $p$ to perform \emph{all} its obligations, positive and negative. Obligation to not perform action $a$ ($\OPoo{p}{!a}$) can be similarly expressed. We combine all positive and negative obligations in the following definition:
\[\begin{array}{l}
\satO{p}{(q_1,q_2)_{q_{\mathcal A}} \goesto{A} (q'_1,q'_2)_{q_{\mathcal A}'}} \df \viable{p}{q_{\mathcal A}}{A}\\
\satO{\other{p}}{(q_1,q_2)_{q_{\mathcal A}}} \df \exists A \in \acts{q_{\other{p}}},\;A' \subseteq G^c \st \viable{p}{q_{\mathcal A}}{A\cup A'}
\end{array}\]

\noindent The satisfaction constraint for transitions is only applicable if $A$ is not an action set performed asynchronously by $\other{p}$. For other parameters, $\satO{p}{X}$ is true.

\noindent\textbf{Example:}
Continuing the previous example, to satisfy $\embox{sat}_{p}^O$, \emph{all} of $p$'s outgoing transitions must be $s$-free and must have an $f$, while \other{p} should offer at least one transition that contains an $f$ and not an $s$. That is, if at a given state \other{p} offers only outgoing transitions labeled $\{f, s\}$ then she is forcing $p$ to an $s$ in order to have an $f$, and thus not satisfying its part in $p$'s obligations.


%
%
%
%

\paragraph{\textbf{General contract satisfaction}.} It is defined as: $\sat{p}{X} \df \satP{p}{X} \land \satO{p}{X}$.
%
%
%
Based on this, we can now define correctness of a regulated two-party system.

\begin{definition}
A party $p$ is said to be incapable of breaching a contract in a regulated two-party system $R=\langle S_1,S_2\rangle^{\mathcal A}_G$, written $\bi{p}{R}$, if $p$ cannot be in violation in any of the reachable states and transitions of $R$.
%
\end{definition}

Note that a party being breach-incapable is stronger than just being compliant for one specific run --- $\bi{p}{R}$ means that there is no possible trace of $R$, in which $p$ breaches the contract.  

\subsection{Other Modalities}

\begin{definition}
Permissions and obligations are duals under a notion of norm opposites and action absence. We define the opposite of permission and obligation $!\OPpp{p}{x}$ and $!\OPoo{p}{x}$ syntactically in the following manner:

\begin{itemize}
\item Party $p$ not being permitted to perform an action is equivalent to $p$ being obliged not to perform the action:
$!\OPpp{p}{a} \df \OPoo{p}{!a} \quad \quad !\OPpp{p}{!a} \df \OPoo{p}{a}$

\item Party $p$ not being obliged to perform an action is equivalent to $p$ being permitted not to perform the action:
$!\OPoo{p}{a} \df \OPpp{p}{!a} \quad \quad !\OPoo{p}{!a} \df \OPpp{p}{a}$
\end{itemize}
\end{definition}

It should be noted that we are equating lack of permission to do $a$ to an obligation to perform an action set which does not include $a$. Although this seems to go against the intuitive idea of letting a party do nothing as a way of not violating lack of permission, note that (i) since transitions carry sets of actions, the empty set of actions is a way of satisfying the obligation; and (ii) well-formedness (see \refdef{reg_two_party_system}) of the parties ensures that progress is always possible thus making the formulation of lack of permission conform to our expectations.

It is interesting to note that in a two party system there are alternative notions of opposites to permission and obligation. Consider party $p$ not being permitted to perform action $a$. Apart from the interpretation we gave, in which the norm places the onus on party $p$ not to perform $a$, an alternative view is to push the responsibility to $\other{p}$ and interpret it as: \emph{party $\other{p}$ may not provide a viable action set which includes $a$}. This is distinct from $!\OPpp{p}{a}$ (and indeed from the other modalities we have). Similarly, consider party $p$ not being obliged to perform action $a$. The interpretation we adopted permits party $p$ to not perform $a$, but once again, alternative definitions may be adopted. One possibility is to push the responsibility to $\other{p}$ and interpret it as: \emph{party $\other{p}$ must provide a viable transition which does not include $a$}.  These duals, in which the outer negation of a norm also corresponds to shifting of responsibility give an interesting alternative view of norm opposites in a two-party system. Another interesting alternative would be to interpret these negations as modalities whose only effect is the cancelling of existing clauses.  We will not explore these alternative modalities any further in this paper, since the modalities we adopt
provide a clean notion of conflicts, as discussed in Section \ref{sec:conflicts}. Should they be needed for a particular application, any of the above mentioned interpretations could be included as alternative type of negation. One of the advantages of clear formal semantics is that there is no need to dispute the meaning of a given term, since different ones can be defined and the appropriate one be picked to convey specific meanings.  Prohibition can now be defined as the dual of permission:

\begin{definition}
Prohibition contract clauses $\OPff{p}{a}$ and $\OPff{p}{!a}$, prohibiting party $p$ from performing and not performing $a$ respectively, can be expressed in terms of permission:\\
\centerline{$\OPff{p}{a}~\df~!\OPpp{p}{a} \quad \quad \OPff{p}{!a}~\df~!\OPpp{p}{!a}$}
\end{definition}

\noindent These definitions allow us to express prohibition in terms of obligation not to perform an action:

\begin{proposition}
Prohibition to perform an action is equivalent to obligation not to perform the action: $\OPff{p}{x}=\OPoo{p}{!x}$.
\end{proposition}

\subsection{Contract Strength}
\label{sec:strictness}

\noindent We can now define strictness relationships over contracts.

\begin{definition}\label{def:bi}
A contract automaton ${\mathcal A'}$ is said to be \emph{stricter than} contract automaton ${\mathcal A}$ for party $p$ (or ${\mathcal A}$ said to be \emph{more lenient} than ${\mathcal A'}$ for party $p$), written ${\mathcal A} \LEQcontract_p {\mathcal A'}$, if for any systems $S_1$ and $S_2$, $\bi{p}{\langle S_1,S_2\rangle^{{\mathcal A'}}_G} \implies \bi{p}{\langle S_1,S_2\rangle^{{\mathcal A}}_G}$. We say that two contract automata ${\mathcal A}$ and ${\mathcal A'}$ are equivalent for party $p$, written ${\mathcal A}=_p {\mathcal A'}$, if ${\mathcal A} \LEQcontract_p {\mathcal A'}$ and ${\mathcal A'} \LEQcontract_p {\mathcal A}$. We define global contract strictness ${\mathcal A} \LEQcontract {\mathcal A'}$ to hold if ${\mathcal A} \LEQcontract_p {\mathcal A'}$ holds for all parties $p$, and similarly global contract equivalence ${\mathcal A} = {\mathcal A'}$.
\end{definition}

\begin{proposition}
The relation over contracts $\LEQcontract$ is a partial order. 
\end{proposition}

\noindent Structurally isomorphic contract automata provide a useful proof technique:

\begin{proposition}
\label{prop:isomorphic_ca}
Given two structurally isomorphic contract automata ${\mathcal A}$ and ${\mathcal A}'$, ${\mathcal A} \LEQcontract {\mathcal A}'$ if and only if, for any state or transition $X$, $\satA{{\mathcal A}'}{p}{X} \implies \satA{{\mathcal A}}{p}{X}$.
\end{proposition}

\begin{TR}
This proof principle can be proved to hold by showing that (i) the automata obtained from the synchronous composition with the two contracts are structurally identical; and (ii) using the definition of breach incapability. The principle can be used to prove that contract automata are monotonic:
\end{TR}
\begin{PAPER}
\noindent The full proof of the proposition can be found in~\cite{PS_FLACOS2012-TR}.
\end{PAPER}

\begin{proposition}
Contract automata are monotonic: given two structurally isomorphic contract automata ${\mathcal A}$ and ${\mathcal A}'$, with contract clause functions $\contractF$ and $\contractF'$ respectively, which satisfy that $\forall q \st \contractF(q) \subseteq \contractF'(q)$, it follows that ${\mathcal A}\LEQcontract {\mathcal A}'$.
\end{proposition}

\begin{TR}
The proof follows from the observation that $\sat{p}{X}$ is essentially a conjunction of a proposition for each contract clause in the state. Hence, $\satA{{\mathcal A'}}{p}{X}$ (which has a larger set of clauses) implies $\satA{{\mathcal A}}{p}{X}$. Applying Proposition \ref{prop:isomorphic_ca} to this observation completes the proof.
\end{TR}

Although contracts are expressed as automata, we would like to be able to compare individual clauses. To do this we will need to relate contract automata which are equivalent except for a particular clause replaced by another. 

\begin{definition}
Given two contract clauses $C$ and $C'$, the relation over contract automata $[C \rightarrow C'] \subseteq \CA \times \CA$ relates two contract automata ${\mathcal A}$ and ${\mathcal A'}$ if ${\mathcal A}$ is equivalent to ${\mathcal A'}$ except possibly for a number of instances of clause $C$ replaced by $C'$.

We extend the notion of strictness to contract clauses. We say that clause $C'$ is stricter than clause $C$ for party $p$, written $C \LEQcontract_p C'$, if for any contract automata ${\mathcal A}$ and ${\mathcal A'}$ such that $({\mathcal A},{\mathcal A'}) \in[C \rightarrow C']$, it follows that ${\mathcal A} \LEQcontract_p {\mathcal A'}$. We similarly extend the notion of strictness for all parties $\LEQcontract$.
\end{definition}

The following proposition allows us to use the proof principle given in Proposition \ref{prop:isomorphic_ca} for reasoning about clause strictness:

\begin{proposition}
\label{prop:replace_isomorphic}
Given clauses $C$ and $C'$, any two contract automata related by $[C \rightarrow C']$ are structurally isomorphic.
\end{proposition}

%
%

\subsection{Strictness Theorems}

The strictness relationship between clauses allows us to state the following theorems.

\begin{theorem}
\label{th:pp_p_oo_p}
Obligation is stricter than permission: (i) $\pp_p(a) \LEQcontract \oo_p(a)$; and (ii) $\pp_p(!a) \LEQcontract \oo_p(!a)$.

\begin{proof}
We present the proof of (i) --- the proof of (ii) is very similar. We need to prove that for any contract automata ${\mathcal A}$ and ${\mathcal A}'$ such that $({\mathcal A}, {\mathcal A}') \in [\pp_p(a) \rightarrow \oo_{p}(a)]$, then it follows that ${\mathcal A} \LEQcontract {\mathcal A}'$. Using Proposition \ref{prop:replace_isomorphic}, we know that ${\mathcal A}$ and ${\mathcal A}'$ are structurally isomorphic, allowing us to apply the proof principle of Proposition \ref{prop:isomorphic_ca}.

We thus have to show that $\satA{{\mathcal A'}}{p}{X}$ implies $\satA{{\mathcal A}}{p}{X}$. Since the permission in ${\mathcal A}$ which is replaced by an obligation, never yields violations for party $p$ nor for any party on transitions, it suffices to prove that this implication holds on states for party $\other{p}$.

\noindent The satisfaction function for \other{p}'s obligations in states is:\\
\centerline{$\exists A \in \acts{q_{\other{p}}},\;A' \subseteq G^c \st \viable{p}{q_{\mathcal A'}}{A\cup A'}$}\\
If $a \in G$, and since $a\in O_p(q_{\mathcal A'})$, we can conclude that $a \in A$:\\
\centerline{$a \in G \implies
  \exists A \in \acts{q_p},\;A' \subseteq G^c \st 
    a \in A \land \viable{\other{p}}{q_{\mathcal A'}}{A\cup A'}$}\\
Furthermore, since $q_{\mathcal A}$ has less obligations than $q_{\mathcal A'}$, viability for $q_{\mathcal A'}$ implies viability for $q_{\mathcal A}$:\\
\centerline{$a \in G \implies
  \exists A \in \acts{q_p},\;A' \subseteq G^c \st 
    a \in A \land \viable{\other{p}}{q_{\mathcal A}}{A\cup A'}$}\\
Hence, the satisfaction function for the permission $\pp_p(a)$ holds and thus, by Proposition \ref{prop:isomorphic_ca} we can conclude that ${\mathcal A} \LEQcontract {\mathcal A}'$.

\end{proof}
\end{theorem}

\begin{theorem}
\label{th:pp_p_oo_other_p}
For synchronised actions, obligation for one party is stricter than permission for the other: (i) $\pp_p(a) \LEQcontract \oo_{\other{p}}(a)$; and (ii) $\pp_p(!a) \LEQcontract \oo_{\other{p}}(!a)$.

\begin{TR}
\begin{proof}
As in the previous theorem, we observe that $\pp_p(a)$ can only yield violations for states and for party $\other{p}$.

Observe that the obligation $\OPoo{\other{p}}{a}$ in a state $q_{\mathcal A'}$ guarantees that all outgoing transitions from the state $(q_1,q_2)_{q_{\mathcal A'}} \goesto{A} (q'_1,q'_2)_{q'_{{\mathcal A}'}}$ satisfy $\viable{\other{p}}{q_{\mathcal A'}}{A}$.

Since we assume that the system does not deadlock, there is at least one such transition which party $p$ participates in. Furthermore, if $a\in G$, it must also appear in the actions on the transition:
$$
a \in G \implies
  \exists A \in \acts{q_p},\;A' \subseteq G^c \st 
    a \in A \land \viable{\other{p}}{q_{\mathcal A'}}{A\cup A'}
$$
This guarantees that $\satSingle{(q_1,q_2)_{q_{\mathcal A}}}{p}{\OPpp{p}{a}}$, and allows us to complete the proof using Proposition \ref{prop:isomorphic_ca}.
\end{proof}
\end{TR}
\end{theorem}

It is interesting to note that if we had a weaker semantics which simply identifies a violation without identifying the guilty party, we would be able to show equivalence between $\oo_p(a)$ and $\oo_{\other{p}}(a)$, since a lack of $a$ on a transition would cause a violation of both obligations. However, since our semantics characterise violations for the parties separately, and the partial order $\LEQcontract_p$ is parametrised by the party, we can show that the two obligations are in fact different~\cite{PS_JURIX2011}.

\subsection{Mutually Exclusive Actions}
Although we adopt a multi-action approach, modelling real-world scenarios means that certain actions should never occur concurrently. For instance, one would expect that the automata never perform the action \emph{openDoor} and \emph{closeDoor} on the same transition. This allows us to identify strictness laws which hold only for mutually exclusive actions.

\begin{definition}
Given a multi-action automaton $\langle \Sigma,\;Q,\;q0,\;\rightarrow\rangle$, two actions $a$ and $b$ ($\{a,b\}\subseteq \Sigma$) are said to be mutually exclusive, written $a \mexcl b$, if they can never appear in the same set of actions on transitions. Thus, for any automaton, it should be the case that:\\
\centerline{$\forall (q,A,q') \in \rightarrow \st a \in A \implies b \notin A$}
\end{definition}

\noindent In the rest of the article we will assume that mutually exclusive actions never appear in the synchronisation sets.
\begin{PAPER}
Removing this restriction, however, does not affect the results we present.
\end{PAPER}
\begin{TR}
This is done to simplify the presentation, since otherwise we would need a more complex rule for synchronous composition (not allowing synchronisation when the asynchronous actions of party are in conflict with those of the other) and a modified definition for the satisfaction of obligations (the other party must provide a viable action set which does not include any actions which may conflict with the obligations of the party to whom the obligation applies). Removing this restriction, however, does not affect the results we present.
\end{TR}
The following theorem shows how mutually exclusive actions and action absence are related together under both obligation and permission:

\begin{theorem}
\label{th:mexcl_same}
If $a \mexcl b$ then (i) $\OPoo{p}{!a} \LEQcontract \OPoo{p}{b}$; and (ii) $\OPpp{p}{!a} \LEQcontract \OPpp{p}{b}$.
\begin{TR}
\begin{proof}
To show (i), we need to prove that for any contract automata ${\mathcal A}$ and ${\mathcal A}'$ such that $({\mathcal A}$, ${\mathcal A}') \in [\oo_p(!a) \rightarrow \oo_p(b)]$, then it follows that ${\mathcal A} \LEQcontract {\mathcal A}'$. As in the previous proofs, we can use Proposition \ref{prop:replace_isomorphic} to conclude that ${\mathcal A}$ and ${\mathcal A}'$ are structurally isomorphic, allowing us to apply the proof principle of Proposition \ref{prop:isomorphic_ca}.

We thus have to show that $\satA{{\mathcal A'}}{p}{X}$ implies $\satA{{\mathcal A}}{p}{X}$. We look at transitions and states separately:

\begin{description}
\item[Transitions:]
The satisfaction function for the combined obligations for a transition $(q_1,q_2)_{q_{\mathcal A'}} \goesto{A} (q'_1,q'_2)_{q'_{\mathcal A'}}$ in automaton ${\mathcal A'}$ is that $\viable{p}{q_{\mathcal A'}}{A}$.  By definition of viability and the obligation $\oo_p(b)$ in $q_{\mathcal A'}$, we can thus conclude that $b\in A$.  However, since $a \mexcl b$, this means that $a\notin A$, from which we can conclude that $\viable{p}{q_{\mathcal A}}{A}$ and hence that the satisfaction function also holds for transitions in ${\mathcal A}$.

\item[States:] 
The satisfaction function applied to states acts on the other party $\other{p}$. For state $(q_1,q_2)_{q_{\mathcal A'}}$, it is defined to be $\exists A \in \acts{q_{\other{p}}},\;A' \subseteq G^c \st \viable{p}{q_{\mathcal A'}}{A\cup A'}$. Since $a\in G$, the proof is identical to the previous case. 
\end{description}

Hence, the satisfaction function for $\oo_p(a)$ holds and thus, by Proposition \ref{prop:isomorphic_ca} we can conclude that ${\mathcal A} \LEQcontract {\mathcal A}'$ and hence (i) holds.

The proof of (ii) follows similarly.
\end{proof}
\end{TR}
\end{theorem}

\noindent A similar result can be shown, but referring to the other party in the contract:

\begin{theorem}
\label{th:mexcl_diff}
If $a \mexcl b$ then $\OPoo{\other{p}}{!b} \LEQcontract \OPoo{p}{a}$.

\begin{TR}
\begin{proof}
We take an approach identical to the previous theorems and prove that for any contract automata ${\mathcal A}$ and ${\mathcal A}'$ such that $({\mathcal A}$, ${\mathcal A}') \in [\oo_{\other{p}}(!b) \rightarrow \oo_p(a)]$, then it follows that ${\mathcal A} \LEQcontract {\mathcal A}'$. Propositions \ref{prop:replace_isomorphic} and \ref{prop:isomorphic_ca} can then be used to complete the proof. As before, we consider the satisfaction relation on states and transitions separately:

\begin{description}
\item[Transitions:]
The satisfaction function for the combined obligations for a transition $(q_1,q_2)_{q_{\mathcal A'}} \goesto{A} (q'_1,q'_2)_{q'_{\mathcal A'}}$ in automaton ${\mathcal A'}$ is that $\viable{p}{q_{\mathcal A'}}{A}$.  By definition of viability and the obligation $\oo_p(a)$ in $q_{\mathcal A'}$, we can thus conclude that $a\in A$.  However, since $a \mexcl b$, this means that $b\notin A$. The same transition must be viable for \other{p} in ${\mathcal A'}$, so $\viable{\other{p}}{q_{\mathcal A'}}{A}$ holds. The absence of $b$ also allows us to conclude that $\viable{\other{p}}{q_{\mathcal A}}{A}$, which is the satisfaction function for \OPoo{\other{p}}{!b} over transitions in ${\mathcal A}$.

\item[States:] 
For state $(q_1,q_2)_{q_{\mathcal A'}}$, since we assume deadlock freedom and satisfaction of the obligation to perform $a$, we know of the existence of an outgoing transition with action $a$ such that $a\in A$. Since party $p$ is participating in this transition, and $a\in G$, we can conclude that there is a transition viable for $\other{p}$, leaving from $q_p$ and with an action set which includes $a$ and hence not $b$. Propositions \ref{prop:replace_isomorphic} and \ref{prop:isomorphic_ca} can then be conclude that $\exists A \in \acts{q_p},\;A' \subseteq G^c \st \viable{\other{p}}{q_{\mathcal A}}{A\cup A'}$. 
\end{description}

\end{proof}
\end{TR}
\end{theorem}

Although one may be tempted to induce that a similar result can be shown for permission (analogous to part (ii) of Theorem \ref{th:mexcl_same}) ---  $\OPpp{\other{p}}{!b} \LEQcontract \OPpp{p}{a}$ does not always hold. As a simple example of a system satisfying \OPpp{p}{a} but not \OPpp{\other{p}}{!b}, consider party $p$ be able to perform just one transition with action set $\{b\}$, and party \other{p} being able to perform one of two transitions: one with action set $\{a\}$, the other with action set $\{ b\}$. Party $p$ is permitted to perform $a$ but party \other{p} is not permitted to perform $!b$.

\section{Conflicts}
\label{sec:conflicts}

Contract clauses are not always compatible with one another. Many definitions of conflict are possible --- in this article we deal only with one particular class of conflicts which focusses on conflicting norms and mutually exclusive actions, but some interesting issues arise from party interdependence.
As expected, the obligation on a party to perform an action $a$ and the obligation on the same party not to perform the same action can never be satisfied together. Another interesting example is that of \OPpp{p}{!a} and \OPoo{p}{a}. Although one is tempted to intuitively think that having the possibility of doing something other than $a$ does not conflict with the obligation of doing $a$, multi-action semantics in contracts are different: to satisfy the permission party \other{p} must provide $a$-free action sets which allow $p$ to satisfy her obligations, but that requires that they contain $a$.
In this section we axiomatise the notion of conflicts in interacting two-party systems and investigate some consequences.

\begin{definition}
Contract conflicts is a relation between contract clauses $\conflictsymbol \in \Clause \leftrightarrow \Clause$ and is defined to be the least relation satisfying the following axioms:

\noindent\textbf{Axiom 1:}  
Opposite permissions conflict: $\vdash \OPpp{p}{x} \conflicts !\OPpp{p}{x}$.

\noindent\textbf{Axiom 2:}  
Obligation to perform mutually exclusive actions is a conflict: $a \mexcl b \vdash \OPoo{p}{a} \conflicts \OPoo{p}{b}$.

\noindent\textbf{Axiom 3:}  
Conflicts are closed under symmetry: $C \conflicts C' \vdash C' \conflicts C$.

\noindent\textbf{Axiom 4:}  
Conflicts are closed under increased strictness: $C \conflicts C' \land C' \LEQcontract C'' \vdash C \conflicts C''$.
\end{definition}

Although conflicts are only identified for opposing permissions in the axioms, they also arise in opposing obligations, and can be shown to follow from the axioms.

\begin{proposition}
\label{prop:conflict_oox_!oox}
Opposite obligations conflict with each other: $\OPoo{p}{x} \conflicts !\OPoo{p}{x}$.
\begin{TR}
\begin{proof}
The proof uses the definition of negated permission and obligation to derive the desired result:
\[\begin{array}{cl}
         & \mbox{definition of conflict on opposing permissions}\\
\implies & \OPpp{p}{x} \conflicts !\OPpp{p}{x}\\
\implies & \mbox{for some $y$, $x=!y$}\\
         & \OPpp{p}{!y} \conflicts !\OPpp{p}{!y}\\
\implies & \mbox{definition of $!\OPpp{p}{y}$ and $!\OPoo{p}{y}$}\\
         & !\OPoo{p}{y} \conflicts \OPoo{p}{y}\\
\implies & \mbox{symmetry of $\conflictsymbol$}\\
         & \OPoo{p}{y} \conflicts !\OPoo{p}{y}
\end{array}\]
\end{proof}
\end{TR}
\end{proposition}
\begin{TR}
Various other conflicts can be derived from the axioms. The following show conflicts between permissions and obligations and arising from enforcing norms over both the presence and absence of an action.
\end{TR}
\begin{proposition}
\label{prop:conflict_oox}
Obligation to perform an action conflicts with both permission and obligation to not perform it: 
(i) $\OPoo{p}{x} \conflicts \OPpp{p}{!x}$; and 
(ii) $\OPoo{p}{x} \conflicts \OPoo{p}{!x}$.
Obligation to perform an action also conflicts with lack of permission to perform the action:
(iii) $\OPoo{p}{x} \conflicts !\OPpp{p}{x}$.

\begin{TR}
\begin{proof}
By Proposition \ref{prop:conflict_oox_!oox}, we know that $\OPoo{p}{x} \conflicts !\OPoo{p}{x}$, which, by definition of $!\OPoo{p}{x}$ is equivalent to $\OPoo{p}{x} \conflicts \OPpp{p}{!x}$, hence completing the proof for (i).

\noindent By result (i) and $\OPpp{p}{!x} \LEQcontract \OPoo{p}{!x}$, we can use the strictness axiom of conflicts to conclude that (ii) holds: $\OPoo{p}{x} \conflicts \OPoo{p}{!x}$.

\noindent Result (iii) follows directly from the definition of $!\OPpp{p}{x}$ and result (ii).
\end{proof}
\end{TR}
\end{proposition}
\begin{TR}
Finally, we show how making two conflicting contracts stricter does not get rid of the conflict:
\end{TR}
\begin{proposition}
Given two conflicting clauses $C_1 \conflicts C_2$, making the two clauses stricter does not resolve the conflict: if  $C_1 \LEQcontract C_1'$ and $C_2 \LEQcontract C_2'$, then $C_1' \conflicts C_2'$.
\begin{TR}
\begin{proof}
The proof follows by applying axiom of closure under increased strictness twice and the axiom of symmetry.
\end{proof}
\end{TR}
\end{proposition}

\begin{figure}
\begin{center}
\begin{tikzpicture}[
  >=latex,
  node distance=1cm,
  auto,
  every state/.style={rectangle,rounded corners,draw,font=\footnotesize},
  every node/.style={font=\footnotesize},
  /tikz/initial text=,
  bend angle=90,
  ]
\node[state, initial] (A1S1) {$\OPff{j}{\mbox{transfer}}$};
\node[state] (A1S2) [right=of A1S1] {$\OPpp{j}{\mbox{transfer}}$};
\node[state, initial] (A2S1) [right=of A1S2]{$\OPpp{j}{\mbox{login}}$};
\node[state] (A2S2) [right=of A2S1] {$\OPff{j}{\mbox{login}},\;\OPff{j}{\mbox{transfer}}$};

\path[->]
  (A1S1) edge [bend left] node {login} (A1S2)
  (A1S2) edge [bend left] node {logout} (A1S1)
  (A2S1) edge [bend left] node {malicious} (A2S2)
  (A2S2) edge [bend left] node {cleared} (A2S1)
;
\end{tikzpicture}
\end{center}
\caption{Internet banking contracts}
\label{f:example}
\end{figure}
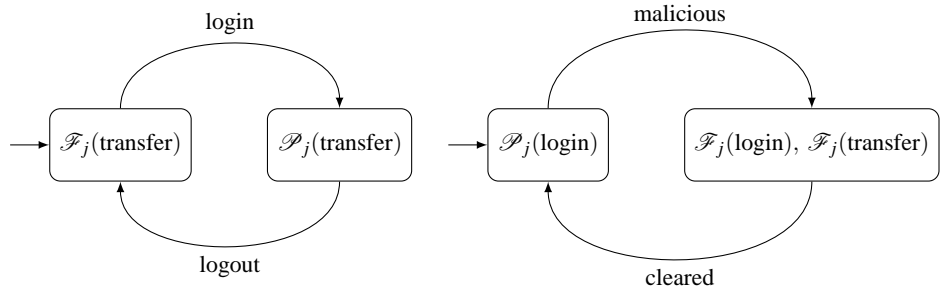

\noindent\textbf{Example:}
As a simple example, consider John signing a contract with his bank. The contract says that (i) whenever he is logged into his Internet banking account, he is to be permitted to make money transfers; and (ii) if a malicious attempt to log in to his account is identified, logging in and making transfers will be prohibited until the situation is cleared. The two statements can be expressed in the two contract automata shown in Fig. \ref{f:example}. Combining the two statements, however results in an automaton where initially, after performing action set $\{\mbox{login},\;\mbox{malicious}\}$, one ends up in a state with both $\OPpp{p}{\mbox{transfer}}$ and $\OPff{p}{\mbox{transfer}}$, which are in conflict.

\section{Related Work}
\label{sec:related_work}

Despite the fact that contracts are, by definition, an agreement between two or more parties, most formal studies regulate the parties independently and do not analyse how permissions, obligations or prohibitions for one party affect the other, or do so in limited ways. Here we summarise the most related work.

\cite{governatori2005dealing} deals with obligation violations in contracts using the domain specific BCL language~\cite{ruleml}, introducing contrary-to-duty clauses and directed obligations, but does not analyse the reciprocity of deontic clauses in a contract. \cite{Marjanovic:2001:TFM:645344.650361} aims at formalisations of contracts for e-commerce but focuses only on analysing temporal consistency.
A related line of research was started by \cite{herrestad1995deontic}, later followed upon by various others (\cite{Tan:1998:LMD:1189859.1189865,Fasli:2001:CRO:646697.703776}, etc.) --- although not explicitly about contracts, they look at a flavour of axiomatic deontic logic with obligations being directed from one individual towards another, termed \emph{directed obligations}. Directed permissions have also been studied, but were termed to be conflicting because of lack of a clear counterparty, following both the \emph{claimant theory} or the \emph{benefit theory}. Once one considers actions that are only realisable by the two parties in synchrony, as our approach does, the concept of permission appears more clearly. Although it does not fully consider many aspects of permission e.g., \OPpp{p}{!a} -- it would be interesting to direct further research to look at the similarities between both approaches, including variations such as \cite{ryu1998specification}.

Our model does not provide explicitly for the notion of \emph{interference} that has been analysed by many, notably 
Hohfeld~\cite{hohfeld1913}
and Lindahl~\cite{lindahl1977position}, It is important to understand, however, that the difference between \emph{vested} and \emph{naked} liberties (i.e., warranty of immunity from interference) relates to a real concern in the context of general law but blurs in the context of a contract where one party allowing the other to perform a shared action, but reserving itself the right to interfere, does not have practical sense. More specifically, in our formal model \OPpp{p}{a} means not only that $p$ may attempt to perform $a$ --- it means that $p$ would succeed in doing $a$ should she try. If the notion of \emph{attempting} to do an action $a$ that can be interfered by others needs to be modeled, then another action $\mathit{attempt\_a}$ should be added and the permission placed onto the latter. Another alternative is to introduce modalities for trying, as in Santos \etal~\cite{santos1997action}.

Lindahl~\cite{lindahl1977position} studies \emph{liberty spaces} to present the concept of \emph{less free than}, a relationship between maximally consistent sets of deontic positions. The general idea is somewhat similar to our definition of strictness; however, as Lindahl notes, most of the maximally consistent sets are incomparable using this relationship, whereas our notion of strictness provides interesting theorems.

Many of the above mentioned authors, and also others, deal with some definition of conflicts but they usually leave out the inconsistencies that arise because of the onuses imposed to the other party (see our example of \OPpp{p}{!a} conflicting with \OPoo{p}{a} in \Sect{conflicts}).

\section{Conclusions}
\label{sec:conclusions}

In this article we extended our formalisation of contracts for interactive systems~\cite{PS_JURIX2011} to deal with absence of actions, mutually exclusive actions and conflicts. The issues raised by interaction between parties, allowing for collaboration and interference, are particularly interesting in the domain of computer-mediated contracts, in which systems typically work in synchrony and proceed only through handshaked actions. Much work has been done in this domain of synchronous systems from a Computer Science perspective, and we believe that our approach allows us to adopt many existing results into the field of contracts. We are currently applying this approach to the analysis of software requirements documents and studying the classes of rights identified in Kanger \emph{et al.} \cite{kanger1966rights} in an interactive setting.

\begin{small}
\bibliographystyle{eptcs}
\bibliography{permissions}
\end{small}

\end{document}